\documentclass[twocolumn]{aa}  
\usepackage{graphicx}
\usepackage{txfonts}
\usepackage{epsfig}
\usepackage{graphicx}
\usepackage{rotating,lscape}
\begin{document}
   \title{Diffuse Light in Hickson Compact Groups: 
The Dynamically Young System HCG 44}

\author{J. A. L. Aguerri\inst{1}, N. Castro-Rodr\'\i guez\inst{1}, N. Napolitano\inst{2} M. Arnaboldi\inst{3,4} \& O. Gerhard\inst{5}}

\offprints{jalfonso@iac.es}

\institute{Instituto de Astrof\'\i sica de Canarias, Calle Via Lactea
  s/n, E-38200 La Laguna, Spain \and Osservatorio Astronomico di Capodimonte, via Moiariello 16, I-80131
  Napoli, Italy. \and INAF, Osservatorio Astronomico di
  Pino Torinese, I-10025 Pino Torinese, Italy \and European Southern
  Observatory, Karl-Schwarzschild-Strasse 2, D-85748 Garching, Germany \and Max-Planck-Institut Institut f\"ur Extraterrestrische
  Physik, Giessenbachstrasse, D-85741 Garching, Germany.}

   \date{\today}

\authorrunning{Aguerri et al.}
\titlerunning{Diffuse Light in Hickson Compact Groups}

\abstract{Compact groups are associations of a few galaxies in which
  the environment plays an important role in galaxy evolution. The low
  group velocity dispersion favors tidal interactions and mergers,
  which may bring stars from galaxies to the diffuse intragroup light.
  Numerical simulations of galaxy clusters in hierarchical cosmologies
  show that the amount of the diffuse light increases with the
  dynamical evolution of the cluster.}{We search for diffuse light in
  the galaxy group HCG 44 in order to determine its luminosity and
  luminosity fraction. Combining with literature data, we aim to
  constrain the dynamical status of Hickson compact groups.}{We use
  Intra Group planetary nebulae (IGPNe) as tracers of diffuse
  light. These are detected by the so-called on band-off band
  technique.}{We found 12 emission line objects in HCG 44, none of
  them associated with the galaxies of the group. The absence of PNe
  in the elliptical galaxy, NGC 3193, implies that this galaxy is
  located behind the group, leaving only three spiral galaxy members
  in HCG 44. 6/12 emission line objects are consistent with being
  IGPNe in HCG 44, but are also consistent with being Ly$\alpha$
  background galaxies. Thus we derive an upper limit to the diffuse
  light fraction in HCG 44 of $4.7\%$, corresponding to 1.06$\times
  10^{9} L_{\odot,B}$ and mean surface brightness of $\mu_{B}=30.04$
  mag arcsec$^{-2}$.  We find a correlation between the fraction of
  elliptical galaxies and the amount of diffuse light in Hickson
  compact groups. Those with large fraction of diffuse light are those
  with large fractions in number and luminosity of E/S0 galaxies. This
  indicates that the diffuse light is mainly created in dynamical
  processes during the formation of bright elliptical galaxies in
  major mergers.}{We propose an evolutionary sequence for Hickson
  compact groups in which the amount of diffuse light increases with
  the dynamical evolution of the group.}


   \maketitle
%

\section{Introduction}

The Hickson Compact Groups (HCG; Hickson 1982) are characterized by a
small number of galaxies (4 to 10) and low velocity dispersion
($\sigma \approx 200 $ km s$^{-1}$) in the group. They have high
galaxy densities (comparable to the cores of galaxy clusters), and are
located in low galaxy density environments. The combination of high
galaxy density in low density environments makes them ideal
laboratories to investigate the role of the environment on galaxy
evolution.  The stability of galaxy groups have been probed by
numerical simulations (G{\' o}mez-Flechoso \&
Dom{\'{\i}}nguez-Tenreiro 2001). Nonetheless, the question is open
regarding their dynamical state.  The fact that we only measure
projected properties of galaxy groups may imply that these groups are
not physically bound systems. Several interpretations were proposed in
the literature: transient dense configurations (Rose 1977; Sargent \&
Turner 1972); chance alignments in loose groups (Mamon 1986, 1995;
Walke \& Mamon 1989); filaments seen end-on (Hernquist et al. 1995).

Galaxies in high density environments may evolve differently than in
the field due to several physical mechanisms affecting their gas
and/or stellar content. Ram pressure stripping (e.g., Gunn \& Gott
1972; Quilis et al. 2000) or starvation (e.g., Larson, Tinsley \&
Caldwell 1980; Bekki et al. 2002) are mechanisms which can affect the
gas content due to the interaction of the galaxies with a hot X-ray
medium. On the other hand, galaxy-galaxy interactions modify the
morphology of the galaxies by stripping galaxy material (gas and
stars), being more efficient when galaxies have low relative
velocities (e.g., Gerhard \& Fall 1983; Mihos 2004). Therefore, it is
likely that tidal interactions modify group galaxies, and may indicate
the evolutionary state of the system. More evolved groups are expected
to be those with galaxies whose morphologies are most disturbed.
  
The search for signs of interactions in group galaxies was driven in
the past by different tracers: tidal features, HI galaxy content, and
X-ray emission. Galaxy-galaxy interactions and galaxy mergers produce
long tidal tails, stripping stars from galaxies which are deposited in
the intragroup region, building-up the intragroup or diffuse
light. Thus, the amount of the diffuse light in groups would be larger
in more evolved galaxy groups. But, tidal stripping can also decrease
the HI content of the galaxies in the groups. In this way,
Verdes-Montenegro et al. (2001) proposed and evolutionary scenario in
which the amount of HI would decrease further with evolution. The
X-ray emission in some Hickson groups has been detected by the ROSAT
satellite (Ebeling et al. 1994; Pildis et al. 1995a), and demonstrates
that the galaxies in X-ray emitting groups are bound and interacting
with a hot intragroup gas.

The presence of diffuse light in rich galaxy clusters was first
suspected in Zwicky's (1951) work on the Coma cluster. It has now been
reliably detected in nearby and rich galaxy clusters in several
different ways: direct detection of intracluster planetary nebulae
(Theuns \& Warren 1997; Arnaboldi et al. 2002, 2003, 2004; Aguerri et
al. 2005; Feldmeier et al. 1998, 2003a, 2004a), detection of
individual intracluster RGB stars (Ferguson et al. 1998; Durrell et
al. 2002); surface brightness measurements after galaxy subtraction
(Zibetti et al. 2005); detection of tidal tails around galaxies in
clusters (Mihos et al. 2005; Adami et al. 2005); deep surface
photometry (Feldmeier et al. 2002, 2004b). These works provide a
consistent estimate that 10-20$\%$ of the light in galaxy clusters is
located in the intracluster region.

Numerical simulations of galaxy cluster formation in $\Lambda$CDM
cosmology show that the origin of the diffuse light is related with
the formation of bright elliptical galaxies by mergers (Murante et al.,
in preparation), and the fraction of ICL depends on the gravitational
matter and the state of evolution of the cluster. Thus, the fraction
of ICL was found to increase from 10-20$\%$ in clusters with $10^{14}
M_{\odot}$ to up to 50$\%$ for very massive clusters with $10^{15}
M_{\odot}$ (Murante et al. 2004). For a fixed mass ($10^{14}
M_{\odot}$), Sommer-Larsen (2006) and Rudick et al. (2006) show that
the fraction of ICL increases also with the degree of dynamical
evolution of the clusters. Those clusters with large differences in
the magnitudes of the two brightest galaxies ($\Delta m >2$) are
evolved clusters and show the largest fraction of the diffuse light.
Some of these predictions have been confirmed in the Virgo cluster
where an important amount of diffuse light have been observed in
regions near to the brightest cluster galaxy M87 and the subgroup
formed by M84 and M86 (Aguerri et al. 2005; Mihos et al. 2005).

Little is known about the diffuse light in groups of galaxies. Pildis
et al. (1995b) studied the diffuse light in 12 galaxy groups, finding
that only one, HCG 94, has diffuse light in the group potential (with
a luminosity of 7$L^{*}$). In contrast, the other groups do not
contain more than $1/3 L^{*}$ in diffuse light. Feldmeier et al.
(2003b) and Castro-Rodriguez et al. (2003) measured the intragroup
light (IGL) by searching for intragroup planetary nebulae (IGPNe) in
the M81 and Leo groups, respectively. They found that at most a few
percent of the light of these groups is located in the intragroup
region. In contrast, White et al. (2003) found that the fraction of
diffuse light in HCG 90 was 38-48$\%$. Da Rocha \& de Oliveia (2005)
also measured the fraction of intragroup light in three Hickson
groups. They found the the IGL contributes 0-45$\%$ to the total
galaxy light. The broad range in the observed diffuse light fraction
was interpreted in the sense that these groups being in different
states of evolution.

The aim of this paper is to study the amount of diffuse light in HCG
44 through the detection of IGPNe, to give a description of the
evolutionary state of the group, and to investigate the implications
of the combined data for the groups analysed so far. The paper is
organized as follows: in Section 2 we present a description of the
group HCG 44. The observations and data reduction are described in
Section 3. The detection of IGPNe in HCG 44 is described in Section 4,
and the fraction of IGL is discussed in Section 5. In Section 6 we
analyze the dependence of the diffuse light fraction on the galaxy
content of compact groups, and the conclusions are given in Section 7.

\section{HCG 44}

HCG 44 is an association of four galaxies: NGC 3190, NGC 3193, NGC
3185 and NGC 3187, located within a circle of 16.4 arcmin diameter and
centered at $\alpha(J2000)=$10:18:05,
$\delta(J2000)=$+21:48:44\footnote{The central coordinates and the
angular diameter of HCG 44 were taken from Hickson (1982)}.  The mean velocity of
the galaxies in the group is 1379 km s$^{-1}$, and the galaxy velocity
dispersion is 219 km s$^{-1}$ (Hickson et al. 1992). This implies a group distance of 18.4 Mpc, assuming $H_{0}=75$ km s$^{-1}$
Mpc$^{-1}$. The distance  to the galaxies of this group was also obtained from the Tully-Fisher (TF) relation and the surface brightness fluctuations method (SBF). Williams et al. (1991) measure the TF distances of the three spiral galaxies in the group (NGC 3190, NGC 3185 and NGC 3187).  They conclude that these three galaxies are at a common distance of 19 Mpc,corresponding to a distance modulus of 31.4. This distance is in good agreement with that obtained from their recessional velocities, and will be the adopted distance to the group in the present paper. Tonry et al. (2001) computed the distance modulus of the elliptical galaxy in the association, NGC 3193, using surface brightness fluctuations. They obtained a distance modulus of 32.71, corresponding a distance of 35 Mpc. Thus, this galaxy is located at a much larger distance than the three spirals and has a large peculiar velocity with respect to the Hubble flow. 

These results show that HCG 44 consists of three spiral galaxies. The brightest galaxy, NGC 3190, is an almost edge on early-type
spiral, showing a strong dust lane and disturbed outer isophotes. Close to NGC 3190 is located the faintest galaxy of the group, NGC 3187. This is a peculiar
barred galaxy with two open arms.  However, Rubin et al. (1991)
proposed that NGC 3187 is not really a barred galaxy, and what we are
observing are two tidal tails coming out of the plane of a spiral
galaxy. They also found strange velocity patterns in these two galaxies, concluding that this is a consequence of a recent tidal interaction. This interaction is also visible in HI: Williams et al. (1991) found a faint HI bridge, connecting NGC 3190 and NGC 3187. This was the only gas detected outwards the galaxies. They conclude that HCG 44 is a  dynamically young group of galaxies. The third spiral galaxy of the group, NGC 3185, does not show signs of interactions. It has a nuclear emission
knot surrounded by a ring of HII regions with no H$\alpha$ emission
within. This galaxy has been reported to be a Seyfert 2 (Huchra \&
Burg 1992). Table 1 shows the main characteristics of the galaxies in HCG
44.

In the present study we will try to give a constraint to the
evolutionary state of the group by measuring the amount of IGL in
HCG 44.

\section{Observations and Data Reduction}
We have observed one field centered on HCG 44 at $\alpha$
(J2000)=10:17:58, $\delta$ (J2000)=+21:48:44 in December  2003,
using the Wide Field Camera (WFC) at the 2.5m Isaac Newton
Telescope (INT) located in La Palma insland. 

The WFC  is an optical  mosaic camera mounted at the prime focus of
the telescope. It consists of 4 thinned EEV 2kx4k CCDs with a pixel
size of 0.33"/pixel. This gives a total field of view of 34$'
\times$34$'$. The galaxy group was imaged through a B band broad
filter and an [OIII] narrow band filter. The broad-band filter
was centered at 4407 $\AA$ and had a width of 1022$\AA$, and the
narrow-band filter was centered at 5027 
$\AA$ and had a width of 60 $\AA$. The exposure times
were 8$\times$3000 s and 24$\times$600 s for the [OIII] and B band
images, respectively. The images were obtained under photometric
conditions and the final seeing was 1.5 arcsec in both filter. After
stacking all the images the total effective area was 874.13
arcmin$^{2}$. Figure 1 shows the B band image of the group.

\begin{table}
\caption{Properties of the galaxies in HCG 44. The only elliptical galaxy in the original HCG 44, NGC 3193, is located $\approx 16$ Mpc behind the group of the spiral galaxies (see text).}
\begin{tabular}{ccccc}\hline
Name&  $\alpha$(1950) & $\delta$ (1950)	&  V (km s$^{-1}$) & B mag. \\
\hline 
\hline	  
NGC3190     & 10:15:20.6 & +22:04:54.9 & 1293   & 11.5 \\
NGC3193     & 10:15:39.6 & +22:08:36.8 & 1378   & 11.6 \\
NGC3185     & 10:14:53.3 & +22:56:18.8 & 1218   & 12.5 \\
NGC3187     & 10:15:02.5 & +22:07:25.4 & 1579   & 13.1 \\
\hline
\end{tabular}
\label{Tab:hcg44_prop}
\end{table}

\begin{figure*}
\begin{center}
\mbox{\epsfig{file=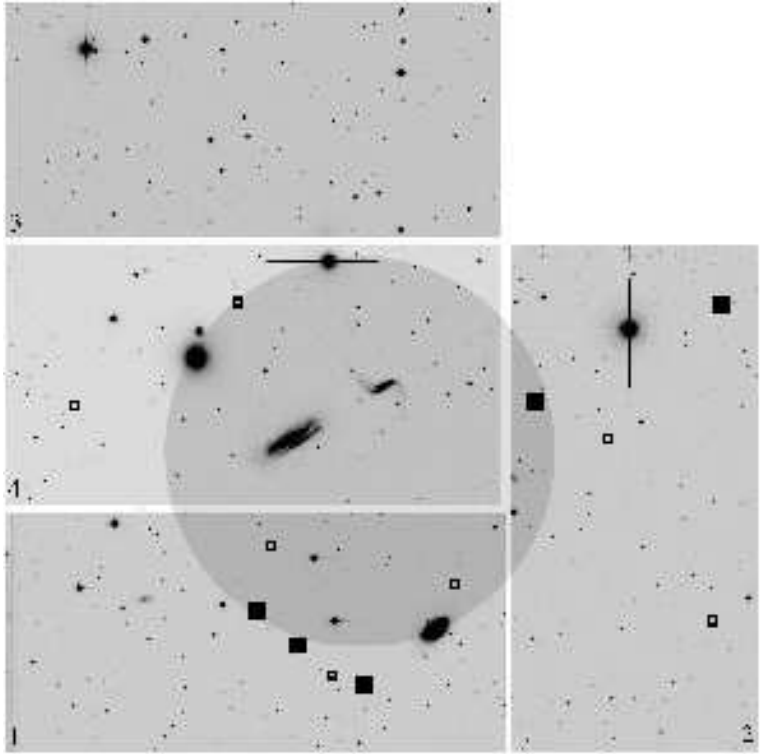}}
\end{center}
\caption{B band image of HCG 44. North is up and East is left. The cross represent the center of the group (Hickson 1982). The darker circle contains the galaxies of the group (see Hickson 1982). The squares show the locations of the [OIII] emission line objects detected in the image. The full squares represent those emission line objects with m(5007)$<26.88$, and open ones are those with $m(5007) \geq 26.88$. The numbers label the CCDs of the WFC (see text for more details).}
\label{Fig:fig1}
\end{figure*}


\subsection{Data reduction}

The data were reduced using the mosaic IRAF tasks package MSCRED. The images were
corrected by dark, bias and flat-field. In addition to the standard flat-fields, we 
created a
superflat image combining the scientific images. Then, the images were
corrected by superflat in order to correct for second order sky structures not
removed by twilight flats, which are usually present in wide
field images. 

The calibration of the broad-band image was obtained using several
Landolt fields. For the narrow-band filter spectrophotometric stars
were observed. Fluxes were then normalized to the AB magnitude system,
following Theuns \& Warren (1997). A detailed description of the
calibration steps and the relation between AB magnitudes and the
``m(5007)'' [OIII] magnitude introduced by Jacoby (1989) are given in
Arnaboldi et al. (2002). For the filters used in the present
observations, the relation between the AB magnitudes and
m(5007) is given by: m(5007)=m(AB)+3.02.

\subsection{PNe as traces of diffuse light in groups and clusters of galaxies}

Arnaboldi et al. (1996) detected for the first time 3 PNe in the Virgo cluster with  different radial velocity from that of the nearby M86 galaxy. This
difference in the radial velocity pointed to the fact that these PNe
were not bound to any galaxy and they were free flying in the Virgo
cluster potential. This demonstrates that the
diffuse light in clusters can be traced by the detection of ICPNe.  
This method has
been used for the Fornax (Theuns \& Warren 1997), several
fields in the Virgo cluster (Arnaboldi et al. 2002, 2003; Aguerri et al.
2005; Feldmeier et al. 1998, 2003a, 2004a), and for several nearby galaxy groups (M81, Feldmeier et
al. 2003b, and Leo, Casto-Rodr\'{\i}guez et al. 2003).

The detection of the ICPNe is based on the strong
emission of these objects in the 5007$\AA$ [OIII] line. PNe are detected using the so-called
ON-OFF band technique (Jacoby et al. 1990). This observational
technique requires two images of the field. One image is taken through a narrow-band filter
(on-band filter) centered at the wavelength of the [OIII]$\lambda 5007
\AA$ emission at the redshift of galaxy group or cluster under study, and
the other image is taken through a broad-band filter (off-band
filter) not containing the [OIII]$\lambda 5007
\AA$ emission. Then, the photometric IGPN 
candidates are those objects detected in the on-band filter and not
detected in off-band filter. They should also be point-like
objects, because PNe at the distance of galaxy clusters or groups can not
be resolved.

\section{Selection of IGPNe candidates}

Our group has developed an automatic procedure for the detection of
 ICPNe candidates in the Virgo cluster (see Arnaboldi et
al. 2002 and Aguerri et al. 2005 for a full description of the method),
which can be applied to HCG 44. This technique is based on the
classification of the detected objects in the on- and off-band images
according to their positions in a color-magnitude diagram (CMD). We
 measure the on- and off-band photometry of all
objects located in the images. This was done using SExtractor (Bertin
\& Arnouts 1996). All objects were plotted in the CMD $m_{n}-m_{b}$ vs
$m_{n}$, being $m_{n}$ and $m_{b}$ the narrow and broad band
magnitudes of each object, respectively.  The most reliable IGPNe
photometric candidates are point-like sources with no detected
continuum emission and observed EW greater than 100 \AA\, after
convolution with the photometric errors as a function of  magnitude
(see Arnaboldi et al. 2002; Aguerri et al. 2005). We selected only objects with EW greater than 100 \AA\ to avoid contamination
from [OII] emitters at $z\approx 0.35$ whose emission falls into the  [OIII] filter for HCG 44, and we must properly take into
account the photometric errors in order to avoid contamination by continuum objects
(see the discussion in Arnaboldi et al. 2002 and Aguerri et al. 2005).

\subsection{Simulations and limiting magnitude of the images}

We run Monte Carlo simulations  to optimize source
detection and determine the [OIII] limiting magnitude of the image of
HCG 44. We distributed several thousands
point-like objects on the scientific images far
away from stars or extended objects. The magnitudes of the simulated
objects were randomly distributed in the range $m_{n}=[14,26]$,
following an exponentially increasing luminosity function (LF). We measured, for
a given detection threshold, how many  simulated objects were
recovered in the photometry, the number of spurious objects due to noise, and the
limiting [OIII] magnitude defined as the magnitude at which 50$\%$ of
the simulated objects were not recovered. This process was repeated
with different SExtractor detection thresholds from 0.7$\sigma$ to
1.3$\sigma$, being $\sigma$ the RMS sky background of the image. The
maximum number of simulated objects and the minimum fraction of
spurious ones were detected with the  1$\sigma$ SExtractor detection
threshold (see Arnaboldi et al. 2002 for more details about this
simulations). Figure 2 shows the CMD for the objects detected in HCG
44 with 1$\sigma$ SExtractor detection threshold. The corresponding
value of the [OIII] limiting magnitude for this detection threshold
is $m_{AB}({\rm OIII})=24.48$.

The B band
limiting magnitude was determined by the expression $m_{B}=-2.5
\log(4\pi\sigma_{{\rm seeing}}^{2}\sigma_{{\rm sky}})+Z_{B}$, where $\sigma_{{\rm seeing}}$ is 
the radius of the seeing disc, $\sigma_{{\rm sky}}$ the rms background of
the B band image, and $Z_{B}$ the zero point of the image. We obtained
$m_{B}=26.04$ for our broad band image. Thus, the broad band image is
$\approx 1.5$ magnitudes deeper than the [OIII] one.

The distance to HCG 44 is 19 Mpc from TF measurements (Williams et
al. 1991), therefore the distance modulus is $31.4\pm0.2$. The brightest PNe
at a distance of 19 Mpc have fluxes of 5.65$\times 10^{-17}$ ergs
s$^{-1}$ cm$^{-2}$ (Ciardullo et al. 2002a) or m(5007)=$26.88^{0.22}_{0.18}$. When
expresed in AB magnitudes, given the characteristics of our filters,
this is equivalent to m(AB)=$23.86^{0.22}_{0.18}$, which is $0.62^{0.18}_{0.22}$ brighter than our
limiting magnitude. So the brightest $\approx 0.5$ magnitude of the
PNLF is accessible to our photometry.

We found that the total number of emission line objects which fulfil or the
conditions for being IGPNe in HCG 44 were 12. Table 2 gives
the coordinates and magnitudes of these objects.

\begin{figure}
\begin{center}
\mbox{\epsfig{file=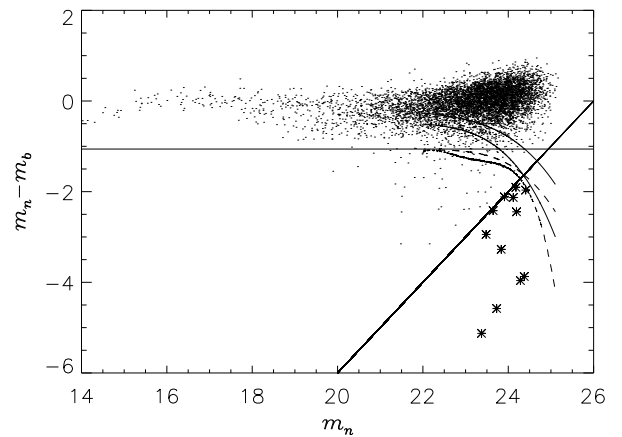,width=9cm}}
\end{center}
\caption{Color magnitude plot of the objects detected by SExtractor
  (small points) in HCG 44. The
  horizontal line indicates objects with an observed EW=100 \AA\. The
  diagonal full line shows the magnitude corresponding to the B band
  limiting magnitude. The full curved lines represent the 99$\%$ and
  99.9$\%$ lines for the distribution of modeloed continuum
  objects. The dashed lines represent 84$\%$ and 97.5$\%$ lines for
  the distribution of modeled objects with $m_{n}-m_{b}=-1$. The
  asterisks represent the emission line candidates with EW$>100$ \AA\ and no detected continuum.}
\label{Fig:fig1}
\end{figure}

\subsection{Background object contamination}

In Aguerri et al. (2005) we carried out a detailed study of the possible
background contaminants in such surveys, mainly: continuum
objects and Ly$\alpha$ background galaxies.

The selection of the candidates was based on a threshold in their
[OIII] fluxes. This can produce a contamination of the sample by
mis-classified faint continuum objects. In Aguerri et al. (2005) this contamination was analyzed, and it was pointed out that this 
contamination is negligible if the off-band image is deep enough. The
difference in the limiting magnitudes between the off- and on- band
images for HCG 44 is 1.5, which ensures us to be free of these
contaminants. This was confirmed by simulations of point-like continuum objects,
similar to Aguerri et al. (2005), which shows that we do not have faint continuum objects in our sample.

The other main contamination comes from Ly$\alpha$ background
galaxies. The  Ly$\alpha$ emission of these objects at $z\approx3$
falls near the [OIII] line at the redshift of HCG 44. We can take into
account this contamination by computing the number of expected
Ly$\alpha$ objects from blank fields in which the emission line
objects were detected using the same technique  (Castro-Rodriguez
et al. 2003; Ciardullo et al. 2002a) . Taking into account the
differences in area, filters width and limiting magnitude we
expect that  the number of Ly$\alpha$ in
HCG 44 should be 25 and 11, respectively. This suggests that all the emission line
objects detected in HCG 44 could be Ly$\alpha$
background galaxies.

\begin{figure}
\begin{center}
\mbox{\epsfig{file=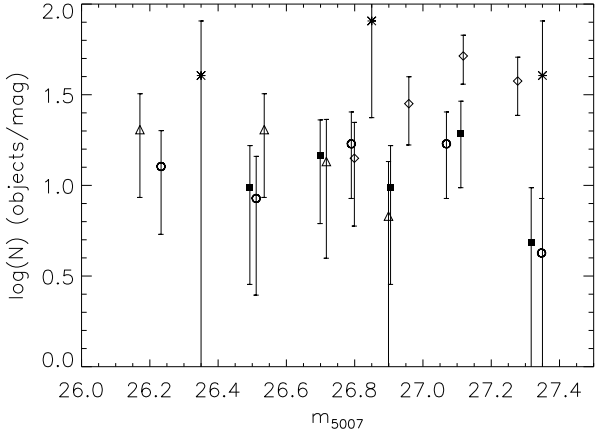,width=9cm}}
\end{center}
\caption{LF of the IGPNe photometrical candidates in HCG 44 (full
  squares). Is is also overplotted the LF of Ly$\alpha$ objects from
  different surveys: (diamonds) Castro-Rodr\'{\i}guez et al. (2003),
  (open circles) LPC field fron Aguerri et al. (2005), (asterisks)
  Kudritzki et al. (2000), and (triangles) Ciardullo et al. (2002a)}
\label{Fig:fig1}
\end{figure}

We have also compared the LF of the emission line objects detected in
HCG 44, those from Ly$\alpha$ surveys (Kudritzki
et al. 2000, Ciardullo et al. 2002a; Castro-Rodriguez et al. 2003), and in one field in the Virgo
cluster (LPC field) which turns out to be compatible with  Ly$\alpha$ objects only (see
Aguerri et al. 2005). The result is shown in Fig. 3. It can be seen
that the LF of the emission line objects detected in HCG 44 is
compatible with the LF of Ly$\alpha$ galaxies. 

\begin{table}[!]
\caption{Emission line objects detected in HCG 44.}
\begin{tabular}{ccccc}\hline
  Name & $\alpha (J2000)$ & $\delta (J2000)$ & m(5007) & m$_{B}$  \\
\hline 
\hline	  
HCG44-ES1 & 10:17:58.0  & +21:39:01.9   &  27.21  & 26.64\\
HCG44-ES2 & 10:18:05.4  & +21:40:24.9   &  26.75  & 28.31 \\
HCG44-ES3 & 10:18:13.4  & +21:41:58.8   &  26.50  & 26.43 \\
HCG44-ES4 & 10:17:33.8  & +21:43:44.4   &  27.43  & 26.37 \\
HCG44-ES5 & 10:18:10.7  & +21:45:03.1   &  27.20  & 26.08 \\
HCG44-ES6 & 10:17:52.8  & +21:38:41.1   &  26.85  & 27.11 \\
HCG44-ES7 & 10:17:04.7  & +21:49:49.1   &  26.93  & 26.03 \\
HCG44-ES8 & 10:17:18.9  & +21:51:29.5   &  26.66  & 26.06 \\
HCG44-ES9 & 10:16:45.0  & +21:41:24.2   &  27.14  & 26.25 \\
HCG44-ES10 & 10:16:42.7  & +21:55:52.9   &  26.39  &    \\
HCG44-ES11 & 10:18:48.2  & +21:51:06.7   &  27.39  & 28.25\\  
HCG44-ES12 & 10:18:17.1  & +21:56:02.1   &  27.30  & 28.24\\  
\hline
\end{tabular}
\label{Tab:candidatos}
\end{table}

\subsection{Are there PNe candidates in individual HCG44 galaxies?}

The detected emission line objects are located in the intragroup
region, away from the galaxies of the group, see Fig. 1. No emission
line objects were detected in those regions where the stellar contiuum
from the group galaxies were also visible. 

How many PNe do we expect to be associated with the HCG44 spiral
galaxies?  The brightest spiral galaxy, NGC~3190, in HCG44 has an
absolute B magnitude $M_B = -19.9$, which is similar to the absolute
magnitude of NGC~5886 and NGC~3351. Ciardullo et al. (2002b) carried
out PNe surveys in these two spirals and the number of PNe detected in
the first 0.5 mag of the PNLF is of order 10; therefore we would
expect similar numbers for NGC~3190. No detection of PNe in NGC~3190
may imply that this galaxy is at a somewhat larger distance than the
19 Mpc derived from the TF relation (Williams et al. 1991). If one
applied the Virgo infall model to the observed redshifts, the mean
distance to NGC 3185 and NGC 3190 would then be 21.7 Mpc
(Kraan-Korteweg 1986), which is slightly larger than the one inferred
from the TF relation.

At the distance of 21.7 Mpc, the bright cut off of the PNe luminosity
function would be at $m(5007)=27.18$, which is 0.32 mag brighter than
our limiting magnitude, and the expected number of PNe would fall by a
factor $>2.0$. A slightly larger distance and the intrinsic dust
absorption in disk galaxies, may be a likely explanation for no-PNe
detected in the spiral galaxies of the HCG44 group.

One of the galaxies of the group (NGC 3193) is an elliptical. At the
distance of HCG 44 the bright cut off of the PNe luminosity function
(PNLF) would be at $m(5007)=26.88$, which is 0.6 magnitudes brighter
than the limiting magnitude of our images. The absolute magnitude of
NGC 3193 is $M_{B}=-19.49$ which corresponds to a luminosity of
$L_{B}=9.55 \times 10^{9} L_{\odot}$.  This means that according to
the luminosity specific PNe density (Ciardullo et al. 1989) for the
M31 bulge for the first 0.5 mag of the PNLF we expect that $\approx
50$ PNe would be detectable in this elliptical galaxy given our
limiting magnitude. The fact that we have not seen PNe in NGC3193
confirms that this galaxy is not at the same distance of the group,
and sets a lower limit of 31.9 to its distance modulus,
consistent with the Tonry et al. (2001) measurement from surface
brightness fluctuations of 32.71. This implies that the bright cut-off
of the PNLF of this galaxy is at $m(5007)=28.21$ which is 0.7
magnitudes fainter than our limiting magnitude. This makes impossible
the detection of PNe associated with this galaxy.

\section{Estimate of fraction of IGL}

We have seen that the emission line objects detected in HCG 44 may be
compatible with Ly$\alpha$ background galaxies, because their
luminosity function is similar to the luminosity function of
Ly$\alpha$ galaxies detected in other surveys. In the Leo group
Castro-Rodriguez et al. (2003) found that the bright cut-off of the LF
of the detected emission line objects was about 1.2 magnitudes fainter
than the bright cut-off of the LFs from PNe associated with the
galaxies of the group. They concluded therefore that the emission line
objects detected in this group were all Ly$\alpha$ background
galaxies. In HCG 44 the PNLF bright cut-off at the group distance is
m(5007)=26.88. Our sample have 5/12 emission line objects brighter
than the PNLF bright cut-off, and 7/12 fainter. Those objects brighter
than the bright cut-off of the PNLF at the group TF distance can not
be PNe. The most probable explanation for them is that they are
Ly$\alpha$ background galaxies or possibly intragroup HII regions (Gerhard et
al. 2002). Nonetheless, the 7 objects fainter than the bright cut-off
may them be IGPNe in HCG 44. We have analyzed the position of these
objects, relative to the group galaxies, and found that 5 of them were
located in CCDs number 1 and 4, among the galaxies of the group, and 2
were located in CCD number 2 (see Fig. 1). One of the objects in CCD
number 2 is located at a distance of 19.6 arcmin from the cluster
center, more than 2 times the radius which encircle the group galaxies
(see Fig 1). The other 6 objects are within 13.6 arcmin of the group
center. Thus, we considered those 6 objects as possible IGPNe, and the
most distant object as a Ly$\alpha$ background object. Then, we can
put an upper limit to the amount of diffuse light in HCG 44 assuming
that the number of IGPNe is 6.

The amount of IGL can be inferred from the stellar light associated
with the IGPNe. This can be computed considering the luminosity
specific planetary nebulae density ($\alpha$). In Aguerri et
al. (2005) we discussed the different possible values of $\alpha$, and
used three representative values, which will be the same adopted in
the present work. For HCG44 the [OIII] limiting
magnitude is 0.62 magnitudes fainter than the PNLF bright cut-off at
the distance of the group. This means that the brightest 0.5 magnitude
of the PNLF is accessible to our photometry. Then, we will use the
value of $\alpha_{0.5,B}$ for computing the stellar light associated
with the IGPNe. But, only 4/6 of the selected emission line objects as
IGPNe are brighter then $M^{*}+0.5$, being $M^{*}$ the PNLF bright
cut-off. We will consider this $4^{+1}_{-0}$ IGPNe to compute the
stellar light associated with them. One value of $\alpha$ corresponds
to that obtained for the bulge of M31 (Ciardullo et al. 1989),
which is $\alpha_{0.5,B}=2.9\times 10^{-9}$.  Then, the luminosity of
the IGL associated to this value in HCG44 is 1.37$\times 10^{9}
L_{\odot,B}$. The second value of $\alpha_{0.5,B}=4.12\times 10^{-9}$
results from the RGB intracluster stellar population observed in Virgo
cluster by Durrell et al. (2002).  For that value, the luminosity of
the IGL in HCG 44 is 0.97$\times 10^{9} L_{\odot,B}$. The third value
of $\alpha$ was taken from the $\alpha$-color relation discovered by
Hui et al. (1993). The mean B-V of the galaxies of HCG 44 is
0.75\footnote{The V magnitudes were taken from de Vaucouleurs et
al. (1991)}. Then, according to the $\alpha$-color relation of Hui et
al. (1993) we have $\alpha_{0.5,B}=4.67\times 10^{-9}$, the luminosity
of the IGL is 0.85$\times 10^{9} L_{\odot,B}$. Taking the mean of
these numbers we give an upper limit of the luminosity of the IGL in
HCG 44 of 1.06$\times 10^{9} L_{\odot,B}$.

We can compare the luminosity of the IGL with the luminosity of the
galaxies of HCG 44. According to their apparent B magnitudes (see
Table 1) the total light from the HCG44 galaxies is 2.15$\times
10^{10} L_{\odot,B}$\footnote{The total luminosity of the galaxies in
the group was computed with NGC 3190, NGC 3185 and NGC 3187. We have
assumed a distance of 19 Mpc for the three galaxies.}; then the upper
limit of the IGL contribution is 4.7$\%$.

We have also computed the surface brightness of the diffuse light in
HCG44. The adopted area is given by the circle centered on the group
center, and with a radius equal to the distance to the center of the
most distant IGPNe in our sample. This corresponds to an area of 581.06
arcmin$^{2}$. Thus, the resulting surface brightness of the diffuse
light in HCG44 is $\mu_{B}=30.04$ mag arcsec$^{-2}$.

The amount of intragroup light depends on the adopted distance
modulus of HCG44. Williams et al. (1991) reported an error of 0.2 to
the measured distance modulus of this group. When this error in the
distance is taken into account, the number of possible IGPNe would be
$4^{+1}_{-0}$. Similar numbers are found when we considered the
distance to the group to be 21.7 $Mpc$ as inferred from the Virgo
infall model to NGC 3185 and NGC 3190. The fraction of diffuse light
in the group is $4.7^{+1.7}_{-0.2} \%$ in both cases.

\section{Implications for diffuse light and dynamical evolution in Hickson Compact Groups.}

Diffuse light has so far always been observed in gravitationally bound
systems such as galaxy groups or clusters. Numerical simulations of
galaxy clusters in hierarchical cosmologies predict that the amount of
diffuse light depends directly on the cluster mass (Murante et
al. 2004), and the dynamical status (Sommer-Larsen 2006; Rudick et
al. 2006). Simulations also indicate that most of the diffuse light
is created during interactions and major mergers resulting in the
largest cluster dominant elliptical galaxies (Murante et al., in
preparation). These predictions are consistent with the observations
in the Virgo cluster where important amounts of diffuse light have
been observed in regions near the dominant galaxy M87 and the subgroup
formed by M84 and M86 (Aguerri et al. 2005; Mihos et al. 2005).

HCG 44 was catalogued by Hickson (1982) as a galaxy group formed by
four members. However, SBF measurements of the elliptical galaxy, NGC
3193, showed that this galaxy is at a larger distance than the rest of
the group (Tonry et al. 2001). Two of the spiral galaxies, NGC 3190
and NGC3187, show morphological distortions which indicate the
interaction between them. This interaction is also visible through the
faint HI bridge between the two galaxies (Williams et al. 1991).
Pildis et al. (1995a) did not find an extended X-ray emission in HCG
44, Ponman et al. (1996) gave an upper limit of the X-ray emission of
this group, being $L_{X}<7\times 10^{40}$ erg s$^{-1}$. This probably means
that this group does not have a deep potential well. Those results
together with the low upper limit of diffuse light in HCG 44  indicate that HCG 44 
is not a dynamically evolved system, or only the two spiral galaxies
are group members.  This would fit well in the framework of
Diaferio et al. (1994) which suggests that compact groups with low or
no X-ray emission and containing galaxies with signs of interactions
would have formed only recently.

The fact that HCG 44 is not a dynamically evolved system opens a long
posed question about the dynamical state of these galaxy
associations. The Hickson (1982) catalog is based on projected galaxy
properties, and a large fraction of those cataloged as compact groups
may not be physically related galaxy systems or  dynamically
evolved systems. As mentioned before the detection of diffuse light in
clusters is always associated with gravitationally bound galaxy
associations, and its amount depends on the dynamical evolution of the
system (Sommer-Larsen 2006; Rudick et al. 2006). Thus, the detection of
diffuse light in galaxy groups can tell us about the dynamical state
of the system.

Several studies of diffuse light in groups have been carried out in
the past.  Pilis et al. (1995b) studied the diffuse component in 12
Hickson Compact Groups. They found that only one of the groups
contained a significant amount of diffuse light. Feldmeier et al. (2003b) and Castro-Rodr\'{\i}guez et
al. (2003) searched for IGPNe in the M81 and the Leo groups,
respectively. They found that a few percent of the light in these
groups is located in the intragroup regions. In contrast, White et
al. (2003) found that the fraction of diffuse light in HCG 90 was
38-48$\%$. Da Rocha \& de Oliveira (2005) found a broad range (0-46
$\%$) of diffuse light for three compact groups. These results
indicate that only $\approx 25\%$ of the so far studied groups
show considerable amount of diffuse light. Those that they are dynamically bound
and evolved systems.
 
We have studied the Hickson compact groups with measurements of
diffuse light in the literature (HCG 44, HCG 79, HCG 88, HCG 90, HCG
94 and HCG 95).  Their diffuse light components were observed using
different methods, and we briefly summarise the strategies and the results.
 
{\it HGC 44- } The search for diffuse light in HCG44 was carried out by
the detecting IGPNe associated with the stellar light bound to the
gravitational potential of the group but not to individual
galaxies. The result from the current analysis is $4.7 \%$

{\it HCG 94} - this was the only group studied by Pildis et
al. (1995b) with diffuse light. They measured the diffuse light by
analysing the residuals after galaxy subtraction on the images,
obtaining $M_{V}=-23.7$ and $B-V\approx1.05$ (see Pildis et
al. 1995b). Then, the luminosity of diffuse light in this group is
$L_{B}=1.67 \times 10^{11} L_{B,\odot}$, representing 42$\%$ of the
light in the group. 

{\it HCG 79, HCG 88 and HCG 95} - These groups were studied by Da
Rocha \& de Oliveira (2005) using wavelet technique. After assuming a
scale for the physical extentions of group galaxies, the diffuse light
is defined as any additional light which is asymmetric and/or outside
galaxies. One of the groups (HCG 88) contains no diffuse
light. However, the diffuse light component in HCG 79 and HCG 95
represents 46$\%$ and 11$\%$ of the total light, respectively. The
mean surface brightness of the diffuse component is $\mu_{B}=27.3$ and
$24.8$ mag arcsec$^{-2}$ for HCG 95 and HCG 79, respectively. 

{\it HCG 90} - This last group was studied by White et al. (2003), who
considered as diffuse component all the light in structures whose
surface brightness is fainter than $\mu_{V}=22.5$ mag
arcsec$^{-2}$. The light in these structures amounts to 38$\%$ of the
total light in the group.  Given the bright surface brightness
cut-off, some of the luminous structures may indeed be associated with
tidal tails or plumes still bound to group galaxies. A cautionary
approach may consider the amount of diffuse light reported 
for this group only as upper limit.

\begin{figure}
\begin{center}
\mbox{\epsfig{file=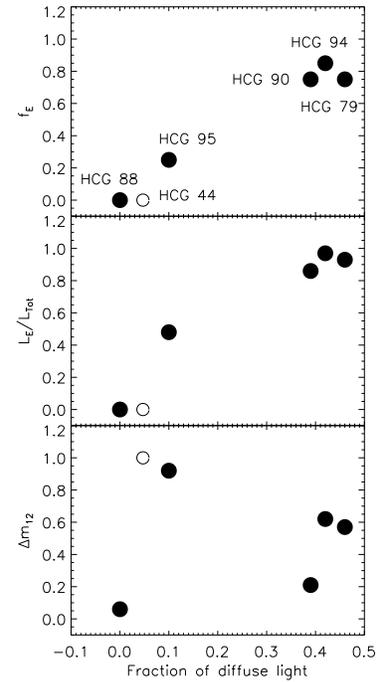,width=13cm}}
\end{center}
\caption{Relation between the amount of diffuse light and fraction of
E/S0 galaxies (top panel), fraction of light from E/S0 galaxies
(middle panel) and magnitude difference between the two brightest
galaxies of the group (bottom panel). The points represent the Hickson
groups: HCG 44, HCG 79, HCG 88, HCG 90, HCG 94 and HCG 95.}
\label{Fig:fig1}
\end{figure}

 We have investigated the relation between the amount of diffuse light
 and their X-ray luminosity and galaxy types. No correlation was found
 between X-ray luminosity and amount of diffuse light, most X-ray
 measurements are only upper limits. However, we found a relation
 between the amount of diffuse light and the fraction of E/S0 galaxies
 (Hickson 1982; Hickson et al. 1989). Figure 4 shows this relation:
 the groups with the lowest fraction of diffuse light are HCG 88 and
 HCG 44, both of them contain no E/S0 (see Hickson 1982). HCG 95 has
 11$\%$ of light as diffuse light (Da Rocha \& de Oliveira 2005) and
 the fraction of E/S0 galaxies for this group is 0.25 (Hickson
 1982). The largest amount of diffuse light were measured for HCG 79,
 HCG 90 and HCG 94, being: 46$\%$ (Da Rocha \& de Oliveira 2005),
 38-40$\%$ (White et al. 2003) and $42\%$ (Pildis et al. 1995b),
 respectively. These are the groups with the largest fraction of
 ellipticals: 0.75, 0.75 and 0.85 for HCG 79, HCG 90 and HCG 94,
 respectively. There is also a relation between the fraction of light
 from E/S0 galaxies in the group (see Fig. 4). Groups for which the
 galaxy light is dominated by E/S0 galaxies (HCG 79, HCG 90 and H94)
 show the largest fraction of diffuse light. We found no correlation
 between the amount of diffuse light and the magnitude difference
 between the two brightest galaxies of the group (see Fig 4), in
 contrast to what is claimed in the numerical simulations of diffuse
 light in $10^{14} M_{\odot}$ galaxy structures (see Sommer-Larsen
 2006). The results in Fig. 4 indicate that the diffuse light is
 created during the formation process of elliptical galaxies.

We can define an evolutionary sequence for galaxy groups according
to the fraction of elliptical galaxies and the amount of diffuse
light. At the beginning of this evolutionary process would be those
groups with small amount of diffuse light and large fraction of spiral
galaxies. The evolution of the groups would continue through
interaction between the galaxies, some of which may lead to a major
merger and, as a final result, an elliptical galaxy. This evolution is
favored by the small velocity dispersion in galaxy groups. The
evolution ending with a major merger, in the group environment, would
also cause mass to be lost from galaxies by tidal processes, which is
deposited in the intragroup region and forms the diffuse
light. Subsequent major mergers will produce more ellipticals and more
diffuse light. At the end of this evolutionary process the group would
be dominated by a single large elliptical galaxy and also will contain
the largest amount of diffuse light, as in the so-called fossil groups
(D'Onghia et al. 2005).

 \section{Conclusions}

 We have studied the presence of IGL in the Hickson compact group HCG 44
 through the detection of IGPNe, using the on-off band
 technique similarly as in the detection of ICPNe in the Virgo cluster. We
 have detected 12 emission line objects. No point-like emission line
 objects have been detected that are associated with the galaxies. 

While the lack of detected emission line objects in the spiral
galaxies of the group could be related to extinction in these
objects, this is not the case for the elliptical galaxy of the group,
NGC 3193. The non-detection of PNe in NGC 3193 sets a lower limit on
its distance modulus, m-M=31.9. This is consistent with the SBF measurement
of (Tonry et al.\ 2001) which places the galaxy at $D= 35$ Mpc, far
behind HCG 44. Thus HCG 44 has effectively only three galaxy members.

Of the 12 emission objects, 6 are fainter than the bright cutoff of
the PNLF and are within the 13.6 arcmin radius from the center of the
group. These objects are compatible with being IGPNe in HCG 44, but
they are also compatible with being Ly$\alpha$ background galaxies,
based on the number density and luminosity function of these objects.
Thus we determine an upper limit to the diffuse light in HCG 44 of
$1.06\times 10^{9} L_{\odot,B}$, corresponding to a mean surface
brightness of $\mu_B=30.04$ mag arcsec$^{-2}$, or 4.7\% of the total
light of the group.

The low upper limit of IGL in HCG44 together with the non-detection of
extended X-ray emission in this group, and the fact that the group
contains only three spiral galaxies, all point towards the
interpretation that this group is a dynamically young system.  The
Hickson catalogue of compact groups includes systems at different
stages of their dynamical evolution. X-ray limits and diffuse light
fractions from the literature indicate than only $\approx 25 \%$ of
the Hickson compact groups are likely to be dynamically evolved
systems.

We have found a correlation between the fraction of elliptical galaxies
and the amount of diffuse light in Hickson compact groups. Those with
large fraction of diffuse light are those with large fractions in
number and luminosity of E/S0 galaxies. No correlation was found
between the amount of diffuse light and the difference in magnitude
between the two brightest galaxies in the group. We suggest an
evolutionary sequence for galaxy groups such that groups with large
fraction of spirals and small amount of diffuse light are dynamically
young systems, and those with large fraction of ellipticals and large
amount of diffuse light are the most evolved. The correlation we found
indicates that the diffuse light is mainly created in dynamical
processes during the formation of bright elliptical galaxies in major
mergers.

\begin{acknowledgements}

We wish to thank to R. Kraan-Korteweg for giving us information about the Virgo infall model for the galaxies of this group. The INT telescope is operated on the island of La Palma by the ING
 Group in the Spanish Observatorio del 
 Roque de Los Muchachos of the Instituto de Astrofísica de
 Canarias. We acknowledge the financial support to JALA and NCR by the
 grant AYA2004-08260. JALA wish also thank the travel support from ESO Director General Discretionary Founds 2006 during the writing process of this manuscript.

\end{acknowledgements}

\end{document}